\documentstyle[preprint,aps]{revtex}
\begin{document}

%\draft

\title{The Temperature Dependence of Solar Neutrino Fluxes}

\author{John N. Bahcall} 
\address{Institute for Advanced Study, Princeton, New Jersey~08540}

\author{Andrew Ulmer}
\address{Princeton University Observatory, Princeton, New Jersey~08544}

\maketitle

\begin{abstract}
By comparing neutrino fluxes and central temperatures
calculated  from 1000 detailed numerical solar models, we 
derive improved scaling laws  which show how each of the neutrino fluxes
depends upon the central temperature (flux $\propto T^m$); we also 
estimate uncertainties  for the temperature  exponents.
With the aid of a one-zone model of the sun, we  derive  
expressions for the temperature exponents of  the neutrino fluxes.
For the most important neutrino fluxes, 
the exponents calculated with the one-zone model agree to within 
20\% or better with the  
exponents extracted from the detailed numerical
models. The one-zone model provides a  physical understanding
of the temperature dependence of the neutrino fluxes.
For the $pp$ neutrino flux, the one-zone model 
explains
the (initially-surprising)
dependence of the flux upon a negative power of the temperature
and  suggests
a new functional dependence. 
This new function makes explicit the strong anti-correlation
between the $^7$Be and $pp$ neutrino fluxes.  The one-zone model also
predicts successfully the average linear relations between  neutrino
fluxes, but cannot predict the appreciable scatter in a $\Delta \phi_i/\phi_i$
versus $\Delta \phi_j/\phi_j$  diagram.
\end{abstract}

\pacs{appropriate pacs numbers}

\section{Introduction }
\label{sec:intro}

Deciphering the solar neutrino problem offers the combined challenge of
understanding the structure of the solar interior and understanding 
the nature of 
neutrino interactions.
The consensus view at present, in part based upon temperature scalings
discussed in this paper, is that 
the measured solar neutrino fluxes reported in the four operating 
experiments  cannot be explained by hypothesizing changes in 
standard solar models (SSMs).  The most plausible explanations, with
the currently available data, require some extension of the
standard model of electroweak interactions
\cite{b90,f91,p91,b93,bl93,w93,c93,h94,c94,c94pr,ber94,b94,pa95,fol95,h95,b95}.

The long-standing
discrepancy between the observed and the predicted neutrino fluxes
has motivated the study of many non-standard solar models, which
are in  most  cases  {\it ad hoc} 
perturbations of  the standard solar model.
For many of the proposed  changes of SSM input parameters
(e.g. nuclear cross sections,
element abundances, and opacities), the predicted neutrino fluxes are
approximately  characterized by a single derived model parameter, 
the central temperature, $T$.
For small variations of input parameters,
the neutrino fluxes and the central temperature of a detailed 
solar model can be  related by a power law of the \hbox{form\cite{bah88}}
\begin{equation}
\label{phin}
\phi \propto T^m,~~~~~m = {d  \log \phi \over d \log T}.
\end{equation}
For the fundamental $pp$ neutrinos, $m \sim  -1$ and for the
important ${\rm ^8B}$ neutrinos, $m \sim +20$.

The temperature dependences indicated in Eq.~(\ref{phin}) are obtained
from precise calculations with complex stellar evolution codes
that solve coupled partial differential equations.  The results of
these calculations are 
well known and have 
been used in many previous analyses of the implications of solar
neutrino experiments.  Our goal is to show how these simple
dependences result from the basic physics of the problem and to what
extent the parameterization in terms of a central temperature is
sufficient to characterize the neutrino fluxes.  The most initially
suprising result of the stellar evolution calculations 
is that the magnitude of the $pp$ neutrino flux dependence depends 
inversely upon the value of the central temperature.  We shall see
that even this result  has a simple, quantitative
explanation in terms of the nuclear
physics of the energy generation process.

The  scaling  with the central solar temperature
can be used to evaluate neutrino fluxes for
small deviations from the standard solar model.
If a model is known to have a slightly
different central temperature than the SSM,
the neutrino fluxes can be estimated
without detailed numerical calculations.
Many classes of non-standard models 
(involving, e.g., 
rapid rotation in the solar interior, some
variations in  nuclear cross sections,
or the existence of a strong magnetic field in the solar core),
reduce the central temperature.
A quantitative 
determination of the reduction in the  central temperature implied by
a specified change in the input physics requires
detailed numerical modeling.
With the aid of the temperature scaling laws, neutrino fluxes from 
non-standard solar models can be investigated over  large
parameter ranges.

Using previously determined scaling laws of neutrino fluxes with
central temperature, $T$\cite{c93,bah88,bah89},
several authors\cite{bl93,c93,h94,c94,c94pr}
have studied  non-standard 
solar  models and have compared the predicted neutrino fluxes with the
available experimental data from the four operating solar neutrino
experiments, Homestake, Kamiokande, GALLEX, and
SAGE\cite{davis93,suzuki95,abdurashitov94,anselmann94}.
These authors\cite{bl93,c93,h94,c94,c94pr} 
show  that it is
impossible to reconcile the  data from the
four operating solar neutrino experiments 
with the neutrino fluxes predicted  by changes in $T$.
They conclude that  non-standard solar models 
which have different central temperatures than the standard model 
are unlikely to solve the solar neutrino problem.

In the applications described above, 
the conclusions depend to some extent upon  the extrapolation of
the temperature scalings to a larger temperature range than was covered
in the original numerical models from which the scaling laws were
derived\cite{bah88}.
We note that Castellani, Degl'Innocenti, and 
Fiorentini\cite{c93} (see also
references\cite{c94,c94pr})  find  good agreement between the
neutrino fluxes calculated from their non-standard, but 
detailed,  solar models
and the fluxes obtained by scaling with respect to the central solar
temperature.  

The analysis in the present paper provides a physical  justification
for the use of the temperature scalings, even for relatively large
changes in the central solar temperature. 
We present improved determinations for the temperature exponents and
estimates for their uncertainties.  We also give exponents for four
minor neutrino fluxes for which temperature dependences were not
previously available.
By construction, the derived scaling
law for the fundamental $pp$ neutrino flux is consistent with 
the observed solar luminosity.

Our results are complementary to the powerful numerical techniques of 
Castellani et al.\cite{c93,c94,c94pr}, who
have shown that several non-standard solar models have a homologous
temperature dependence, and to the insightful physical analysis of
Bludman\cite{bludman96}, who has argued that it is a reasonable
approximation to regard the solar interior as a single region that can
be largely described by a single parameter, the central solar
temperature.
Taken together, these previous investigations provide strong 
motivation for
taking seriously  the predictions of a single-zone solar model.

In what follows, we shall refer frequently to different fusion
reactions in the $pp$ chain.  For convenience, we summarize in 
Table~\ref{ppreaction} the principal reactions in the $pp$ chain.

In Sec.~II of this paper, we report
scaling laws (with uncertainty estimates for the exponents) 
for all the solar neutrino fluxes that were included in the 1000
numerical solar models calculated in the Monte Carlo study  of
Bahcall and Ulrich\cite{bah88}.
We present, in Sec.~III, a simple one-zone model
for the sun which accounts well for the numerically-derived scaling laws.
This model motivates the use of the new functional form
for the temperature dependence of the $pp$ neutrino flux.  We discuss
in Sec.~IV the Fogli-Lisi sum rule\cite{fol95} on the temperature exponents. 
In Sec.~\ref{sec:correlations}, we display the correlations found between
the values of the principal neutrino fluxes 
in the 1000 solar models and show
to what extent the one-zone model can account for these correlations.
Finally, in Sec.~\ref{sec:conclusion} we summarize and discuss our main
conclusions.

\section{Temperature Exponents: Detailed Solar Models}
\label{sec:numerical}

We have derived power-law
scaling relations  with the central solar (model) temperature, $T$,  for the 
neutrino fluxes from the 1000 numerical solar models
of reference \cite{bah88}.
The numerical models are calculated  with a stellar evolution code 
(the same code each time) that
typically uses a few hundred mass zones.  For these precise solar models, the 
principal input parameters,
 (nuclear cross sections and element abundances) were sampled  within
their ranges of uncertainty.
The average dependences upon the central temperature $T$ of
the neutrino fluxes calculated for these 
models are  represented with reasonable accuracy\cite{bah88} by power-law
relations.

Some particle physicists 
have objected to the term `1000 standard solar models' to
describe this collection of numerical solar models on the grounds that 
the same underlying theoretical model (stellar evolution theory and the
standard electroweak model) is used to calculate all of the solar
models.  For their comfort, we have avoided in this paper describing
the set of models as `1000 standard solar models' and instead refer to
them as `1000 detailed solar models' or `1000 numerical solar models.'

In Figs.~\ref{functions} and \ref{powerlaw} and in
Table~\ref{tableexp}, 
we show  temperature exponents 
that were derived by  minimizing the
residuals in  power-law fits of the fluxes versus
$T$.  The power-law representations generally
describe well the dependences of the neutrino fluxes upon the central
temperature of the solar model.

The temperature exponents 
obtained  here by minimizing the residuals for the power
law fits 
are in approximate agreement with the earlier
values\cite{bah88,bah89} obtained 
by  best visual fits (``chi-by-eye'') of power-law temperature
dependences to the distribution of neutrino fluxes. 
The new (previous) temperature exponents of the neutrino
fluxes are: $m(pp) = -1.1 (-1.2)$; 
$m({\rm ^7Be}) = 10$ (8); 
and
$m({\rm ^8B}) = 24$ (18).

For the first time, scaling exponents derived from the Monte Carlo
experiment are given here for the 
less-numerous neutrinos  from
$pep$, $^{13}$N,
$^{15}$O, and $^{17}$F. In previous applications of the scaling laws, it
was necessary to guess the  temperature 
exponents of these four minor neutrino fluxes 
based upon analogies with the  published exponents for the three
dominant neutrino fluxes.

We estimate uncertainties in the values of the scaling exponents 
by determining the indices
twice, once by minimizing the residuals in the best fit to the
neutrino fluxes and once by minimizing the residuals in the best fit
to the central solar temperature.
This method  attributes all of the scatter in the fits to 
either the flux
or to the central temperature; the calculated range 
provides a reasonable  estimate of the plausible range of the 
scaling exponents.

Table~\ref{tableexp} presents the best-estimates for the scaling 
exponents and
their estimated errors.\footnote{The residuals were minimized,
as presented in Figs.~\ref{functions} and \ref{powerlaw}, in
logarithmic space.  For example, the minimization in temperature was
performed on the expression $\Sigma_i\vert\log\ \phi_i - m\ \log\ T
-{\rm const}\vert$, where $\phi_i$ is the set of 1000 solar model
fluxes.  Absolute values were considered to reduce slightly the weight
of outliers.  The inferred temperature exponents are not sensitive to
the precise way the minimization is achieved.  The best-estimate
exponents given in Table~\ref{tableexp} are the average of the
exponents computed by minimizing the residuals in either flux or
temperature.  The uncertainties presented span the range of the two
solutions.}  We also show in Table~\ref{tableexp} 
the semi-analytic results
obtained in the following section using the one-zone solar model.
We choose to represent the $pp$ flux with a new functional form, 
$\phi_{pp} \propto
[1 - 0.08 (T/T_{{\rm SSM}})^m]$, rather
than (as has become conventional)  $\phi_{pp} \propto T^m$, 
for the reasons described in the next section.

\section{Temperature Exponents: One-Zone Model}
\label{sec:analytical}

Using a static one-zone model of the present-day sun,
we derive in this section approximate scaling laws that agree
satisfactorily with the
results obtained by detailed evolutionary calculations of 
numerical solutions of multi-zone solar 
models.
We assume a fixed temperature and matter density
for the one-zone solar model.

With this extremely simplified model, 
we do not expect to obtain accurate
values for the temperature scaling laws.
However, we shall see that 
the temperature exponents that are obtained agree surprisingly well
with the scaling laws derived from the precise evolutionary solar models.

The last column of  Table~\ref{tableexp} gives
the temperature exponents obtained in this section
for a characteristic one-zone central temperature of $T_c = 14 \times 10^6$K.
The derived exponents are not particularly sensitive to the
assumed characteristic central temperature, $T_c$.
Figure~\ref{scalingvstemp} 
shows the dependence of the 
exponents obtained with the one zone model for $T_c$ 
in the range $12 \times 10^{6}$K to $16 \times 10^6$K.

\subsection{$pp$ and $pep$ Temperature Exponents}
\label{subsec:ppTscaling}

In the one-zone approximation,
the measured solar luminosity can be written
\begin{equation}
\label{L}
L_\odot~ = ~ V (\epsilon_{33} R_{33} + \epsilon_{34} R_{34})
~\approx~ V \epsilon_{33} (R_{33} +  R_{34}),
\end{equation}
where $V$ is volume, $R$ is the rate of a nuclear fusion 
reaction per unit 
volume, $R_{33}$
corresponds to the nuclear reaction ${\rm ^3He} + {\rm ^3He}
\longrightarrow {\rm ^4He} ~+~ 2{\rm p}$ (reaction 4 of
Table~\ref{ppreaction}), 
$R_{34}$
corresponds to the reaction 
\hbox{${\rm ^3He} + {\rm ^4He}\longrightarrow {\rm ^7Be} ~+~ \gamma$}
(reaction 5 of Table~\ref{ppreaction}),
and $\epsilon$ is the amount of energy released by
the fusion cycles corrected for neutrino losses.
The rates of both $R_{33}$ and $R_{34}$ increase rapidly with
temperature; $R_{34}$ increases somewhat faster\cite{bah89}.
For the illustrative purposes of the one zone model, the three
percent difference between $\epsilon_{33}$ and $\epsilon_{34}$ has
been neglected in writing Eq. (\ref{L}).  We have also neglected the
small contributions (less than a few percent in standard solar models)
to the total luminosity 
of the fusion reactions associated with the CNO and $^8$B neutrinos.

The flux of $pp$ neutrinos at earth is
\begin{equation}
\label{ppnucrate}
\phi(pp)  = {V \over 4\pi r_\oplus^2}(2 R_{33} + R_{34}).
\end{equation}
The factor of two appears in  Eq.~(\ref{ppnucrate}) 
because two $pp$ neutrinos are produced in
the first branch of the chain;  a $pp$ reaction is required  to make
each of the two $^3$He nuclei.
In the second branch of the $pp$ chain, the alpha particle acts as a 
catalyst and therefore only one
$pp$ reaction is needed. 

Substituting Eq.~(\ref{L}) into Eq.~(\ref{ppnucrate}), one has
\begin{equation}
\label{ppfluxlum}
\phi(pp) ~
\approx ~ {1 \over 4\pi r_\oplus^2}
\left[ {2 L_\odot \over \epsilon_{33}} - V R_{34}\right].
\end{equation}
Since  $R_{34}$ increases with temperature,
the temperature dependence of the $pp$ neutrino flux is negative, as
was found first  in the detailed evolutionary solar model
calculations\cite{bah88}.

The appropriate functional relation is therefore
\begin{equation}
\label{phinp}
\phi(pp)~ = ~b - a T_c^{m^\prime}
\end{equation}
Here $b = 6.5 \times 10^{10} {\rm cm^{-2} s^{-1}}$ and 
\begin{equation}
\label{lnmprime}
m^\prime(pp) = { d \ln\left[ n({\rm ^3He}) n({\rm ^4He}) \langle 3,4 \rangle\right] \over 
d \ln\left[T\right]}.
\end{equation}
We make use of the convenient notation in which the  reaction
cross section times velocity averaged over a Maxwell-Boltzman
distribution is represented, for the interacting nuclei
$i$ and $j$,
by pointed brackets, $\langle i,j \rangle$
(see Eq. 3.9 of reference \cite{bah89}).
Thus the rate of the ${\rm ^3He} + {\rm ^4He}$ reaction is
represented by $R_{34} = n({\rm ^3He})\,n({\rm ^4He})\langle 3,4 \rangle$.

Both, the equilibrium number 
density of $n({\rm ^3He})$ and $\langle 3,4 \rangle$ are strong functions of temperature.
The density can be found\cite{bah89,clayton83} 
by solving the equilibrium rate equations.
At temperatures representative of  the interior of the Sun, a
good approximation for the number density of $^3$He is
\begin{equation}
n({\rm ^3He}) \approx n\rm{(H)} \sqrt{ {\langle 1,1 \rangle \over 2 \langle 3,3 \rangle} }.
\end{equation}
The rate of the $pp$ reaction is written
$R_{1,1} \equiv n({\rm ^1H})^2 \langle 1,1 \rangle/2$, where
$\langle i,j \rangle  \propto T^{-2/3} \exp(-\tau_{i,j})$,
$\tau_{i,j} = 3E_{0,ij}/kT$, and $E_0$ is the most probable energy
of interaction\cite{bah89}.
We note that $\tau$ is proportional to  $T^{-1/3}$.

The value of $m^\prime$ can be calculated from Eq.~(\ref{lnmprime}).

Using this one-zone model, we can estimate the value of $m^\prime$
as well as the value of $m$,
which is traditionally used to describe the temperature dependence
[see Eqs.~(\ref{phin}) and (\ref{phinp})]:
\begin{equation}
\label{np2}
m^\prime(pp) ~\approx~ 11; ~~~~~~~~m(pp) ~\approx~ 
-m^\prime(pp) (aT^{m^\prime}/b)
~\approx~ -0.08 \,m^\prime(pp)
~\approx~ -0.9,
\end{equation}
where we have evaluated the exponents at the one-zone characteristic
central temperature,
$T_c = 
14 \times 10^6$K.
We have also 
taken $aT^{m^\prime}/b \approx R_{34}/2R_{33}$ to be equal to 0.08, as
predicted by detailed numerical solar models or, less precisely,
by the one-zone model.
The preferred form for the temperature dependence of the $pp$ neutrino
flux is therefore
\begin{equation}
\label{preferred}
\phi(pp)~ \propto~ [1 - 0.08 (T_c/T_{c,{\rm SSM}})^{m^\prime}],
\end{equation}
where $T_{c,{\rm SSM}} = 15.64\times10^6$K.

The scaling exponent for the $pp$ neutrino flux that is derived from
the one-zone model agrees reasonably 
well  with the exponent obtained from precise
solar models (see Table~\ref{tableexp}).
The Monte Carlo study of 1000 detailed 
solar models yields $m^\prime =13.0$
($m = -1.1$).  The one-zone model yields 
$m^\prime =11$($m = -0.9$).  The one-zone model underestimates
the exponent of the temperature dependence of the $pp$ neutrino flux
by about 20\%.

The exponent for the  $pep$ neutrino flux can be obtained with the aid
of  the
analysis of the $pp$ neutrino flux given above, since the rate of the $pep$
reaction (reaction 2 of Table~\ref{ppreaction}) 
depends upon  the rate of the $pp$ reaction as\cite{bah69}:
$R(pep) \propto T^{-1/2}R(pp)$.
Therefore, we can write
\begin{equation}
\label{peprate}
\phi(pep) ~\propto~  T_c^{-1/2}(T_c^{m})
= T_c^{-1.4}.
\end{equation}
In the one-zone model, 
the $pep$ neutrino flux, like the $pp$ neutrino flux, scales like a negative 
power
of the central solar temperature,
as is found in the detailed  solar model results. The numerical scaling
derived from the detailed solar models has a rather large uncertainty,
$m(pep) = -2.4 \pm 0.9$, cf. Table~\ref{tableexp}. 
Moreover, the power-law exponent is larger for the $pep$ neutrino
flux than it is for the $pp$ neutrino flux, which follows from the
fact that the $pep$ rate divided by the $pp$ rate is proportional to
the modulus squared of the electron wave function near the two
protons.
The probability density of the electron is inversely proportional to
the electron velocity\cite{bah69}, 
which is itself approximately proportional to $T^{1/2}$.

Although the $pep$ flux could
be written in a form similar to Eq.~(\ref{phinp}) for the $pp$ 
neutrinos, 
we have chosen for simplicity to  represent
this minor component of the solar neutrino flux as a single power of
$T_c$.

Figure~\ref{scalingvstemp} shows that the derived scaling exponents
for the $pp$ and $pep$ neutrino fluxes do
not depend strongly on the  value of the temperature, $T_c$,
which is assumed to characterize the solar interior in the one-zone model.

We conclude this subsection 
by summarizing the main physical  insight.
The often-quoted  dependence of the $pp$ neutrino flux on a negative
power of the temperature ($\phi(pp) \propto T^{-1}$) results from the
fact that as the central 
temperature gets larger, 
an increasing number of the completions of the $pp$ chain
proceed through the ${\rm ^3He} + {\rm ^4He}$
reaction
(reaction 5 of Table~\ref{ppreaction}). A complete fusion reaction of
four protons being converted to an alpha particle via the 
${\rm ^3He} + {\rm ^4He}$
reaction involves the production of  only 
one $pp$ neutrino (cf. Table~\ref{ppreaction}).
On the other hand, a fusion of four protons via 
the ${\rm ^3He} + {\rm ^3He}$ reaction (reaction 4 of
Table~\ref{ppreaction})
produces two $pp$ neutrinos. The  ${\rm ^3He} + {\rm ^3He}$ reaction
predominates at lower temperatures.  Thus the $pp$ flux is larger at
lower central temperatures.

\subsection{$^7$Be and  $^8$B Temperature Exponents}
\label{subsec:be7b8Tscaling}

The dependences of the other neutrino fluxes upon the central solar
temperature can all be derived as simple power laws, 
$\phi \propto T_c^m$.
The power law exponent, $m$, can be obtained from 
the one-zone model.

The flux of $^7$Be neutrinos can be calculated from the rate equation
\begin{equation}
\label{be7rate}
R({\rm ^7Be} + {\rm e}^-) ~\propto~ n(e) n({\rm ^7Be}) \langle e^-, {\rm ^7Be} \rangle ,
\end{equation}
where $R({\rm ^7Be} + {\rm e}^-)$ is the rate at which $^7$Be captures
electrons in the solar interior (\hbox{i.e.,} reaction 6 of
Table~\ref{ppreaction}). 
Here $\langle e^-, {\rm ^7Be} \rangle$  is, as before,
the reaction cross section times velocity averaged over a Maxwell-Boltzman
temperature distribution.
The $^7$Be
electron-capture  reaction is
much faster\cite{bah89} under  solar interior
conditions than  the competing
proton-capture reaction (reaction 8 of Table~\ref{ppreaction}).
The electron-capture  rate is, in equilibrium,  essentially 
equal to the rate of production of
$^7$Be.  Moreover, the production rate of $^7$Be 
is the same as the reaction rate $R_{34}$ that
determines the temperature exponent $m^\prime$ of the
$pp$ reaction [see Eq.~(\ref{ppfluxlum}) and the following
discussion]. 
Therefore, the $^7$Be electron-capture reaction has the same
scaling index,  $m({\rm ^7Be}) = 11$, that was derived 
in Eq.~(\ref{np2}) for  the $pp$ reaction.
Thus
\begin{equation}
\label{be7exponent}
\phi({\rm ^7Be}) ~\propto~ T_c^{11},
\end{equation}
which is  approximately 10\% larger than  the value of 
$m({\rm ^7Be}) = 10$ that is obtained from the detailed
numerical models (see Table~\ref{tableexp}).

The temperature dependence of the $^8$B neutrino flux can be calculated in
a similar manner. The $^8$B neutrino flux results from a rare branch
of the $pp$ chain (reaction 8 of Table~\ref{ppreaction})
in which $^7$Be captures a proton rather than an electron.
Therefore the $^8$B neutrino flux can be written as
\begin{equation}
\label{b8rate}
\phi(^8{\rm B}) ~ \propto~  R({\rm ^7Be} + {\rm e}^-) {\langle {\rm p}, {\rm ^7Be} \rangle \over \langle e^-, {\rm ^7Be} \rangle}.
\end{equation}
Substituting $T_c = 14\times 10^6$K in Eq.~(\ref{b8rate}), we find:
\begin{equation}
\label{b8exponent}
\phi(^8{\rm B}) ~\propto~ T_c^{25}.
\end{equation}
The temperature dependence is much stronger 
for $^8$B neutrinos than for $^7$Be neutrinos
because  the electron capture rate depends only weakly on temperature
(essentially like 
$T^{-1/2}$\cite{bah62}), 
and the proton capture rate increases rapidly with
temperature (like all strong interaction fusion 
rates).

The  derived scaling exponent for $^8$B, $m(^8{\rm B}) = 25$,  is 
fortuitously 
close to the value of $m = 24$ that is obtained by fitting to the detailed
numerical models (cf. Table~\ref{tableexp}).

Figure~\ref{scalingvstemp} shows that the derived scaling exponents
for the $^7$Be and $^8$B neutrino fluxes vary by less than $\pm 10$\%
as the assumed characteristic central temperature $T_c$ varies between $12
\times 10^6$K to $16 \times 10^6$K.

\subsection{CNO Temperature Exponents}

The temperature dependence of the CNO neutrino fluxes can also be
estimated simply.
For the major part of the CNO 
cycle which leads from $^{12}$C to $^{15}$N, the
slowest  reaction  is
$^{14}{\rm N}(p,\gamma)^{15}{\rm O}$\cite{bah89,clayton83}.
Therefore, the neutrino fluxes that are produced in this part of the 
cycle, from $^{13}$N and $^{15}$O,
will both have approximately the same temperature dependence. (Slight 
differences will
occur due, e.g., to non-equilibrium effects not accounted for in our
static one-zone model.)  
The total number of CNO atoms is approximately constant and mostly in the
form of  $^{14}{\rm N}$.

The temperature exponents  for the $^{13}$N and $^{15}$O neutrino fluxes
can be derived from the rate of the $^{14}{\rm N}({\rm p},\gamma)
^{15}{\rm O}$ reaction by calculating  the logarithmic derivative of the
reaction rate with respect to temperature. Thus
\hbox{$m = [\tau(^{14}{\rm N}({\rm p},\gamma)^{15}{\rm O}) ~ - 2]/3$}.
The temperature  dependence 
of the $^{13}$N and $^{15}$O neutrino fluxes is therefore: 
\begin{equation}
\label{n13o15exponent}
\phi(^{13}{\rm N}),~\phi(^{15}{\rm O}) \,~\propto~ T_c^{20}.
\end{equation}
The exponents derived above are in reasonable agreement with the
scaling laws obtained from the (non-equilibrium) detailed solar
models,
which are $m({\rm ^{13}N}) = 24$ and
$m({\rm ^{15}O}) = 27$ (see Table~\ref{tableexp}).

Finally, we calculate the temperature dependence of the rare 
${\rm ^{17}F}$ neutrino flux.
The slowest 
reaction involved in producing the $^{17}$F neutrinos is
$^{16}{\rm O}({\rm p},\gamma)
^{17}{\rm F}$ .
The temperature dependence of the $^{17}$F neutrinos can be calculated
by analogy with the calculation 
for the $^{13}$N and $^{15}$O
neutrinos. In the derivation for the $^{17}$F neutrinos, we consider
the $^{16}{\rm O}({\rm p},\gamma)
^{17}{\rm F}$ reaction
instead of  the $^{14}{\rm N}({\rm p},\gamma)
^{15}{\rm O}$ reaction.
The  scaling law derived in this way is
\begin{equation}
\label{f17exponent}
\phi(^{17}{\rm F}) ~\propto~ T_c^{23}.
\end{equation}
This result is in satisfactory agreement with
the exponent obtained from the 1000 solar models,
which is $m(^{17}{\rm F}) = 28 $
(see Table~\ref{tableexp}).

Figure~\ref{scalingvstemp} shows that the derived scaling exponents
for the CNO  neutrino fluxes vary by less than $\pm 10$\%
as the assumed characteristic central temperature $T_c$ varies between $12
\times 10^6$K to $16 \times 10^6$K.

\section{Sum Rule for Temperature Exponents}
\label{sec:sum}

The luminosity of the sun can be expressed in terms of the individual
neutrino fluxes, $\phi_i$, as follows\cite{c93,fol95,spiro90,dar91}:
\begin{equation}
L_\odot \propto \Sigma_i \epsilon_i \phi_i ,
\label{lphi}
\end{equation}
where for neutrinos from the $pp$ chain
\begin{equation}
\epsilon_i = 13.366\ {\rm MeV} - \langle q_i\rangle .
\label{epsiloneye}
\end{equation}
Here $\langle q_i\rangle$ is the average energy loss to the star from
neutrinos of type $i$; the values of $\epsilon_i$ can be obtained from
Table~3.2 of Ref.~\cite{bah89}.  Equations~(\ref{lphi}) and (\ref{epsiloneye}) assume that all the
nuclear fusion reactions are in equilibrium, which is a reasonably
accurate approximation for all but the CNO neutrinos and, in the
outer region of the solar core, the reactions that produce and destroy
$^3$He.  For the ${\rm ^{13}N}$ and ${\rm
^{15}O}$ neutrinos, Eq.~(\ref{epsiloneye}) does not apply and the
values must be calculated separately from Table~3.3 of
Ref.~\cite{bah89}.  For these neutrinos, $\epsilon ({\rm ^{13}N}) =
3.457$~MeV and $\epsilon ({\rm ^{15}O}) = 21.572$~MeV.
(Equation (18) would also apply 
to CNO neutrinos if they were in complete
equilbirium. )

Fogli and Lisi \cite{fol95} pointed out that Eq.~(\ref{lphi}), when 
combined
with the fact that the present-day solar luminosity is a known constant,
 implies a sum rule on the temperature exponents.  In our notation,
the Fogli-Lisi sum rule is
\begin{equation}
\Sigma_i \epsilon_i m_i \phi_i = 0 .
\label{sumrule}
\end{equation}

The Monte Carlo exponents given in Table~\ref{tableexp} satisfy the 
Fogli-Lisi sum rule to an accuracy of 5\% or better
(i.e., $
{(\Sigma_i \epsilon_i m_i \phi_i)}/
{(\Sigma_i \epsilon_i  \phi_i)}$ is less than 5\%) .
For the one-zone
model, the sum rule is only satisfied to an accuracy of $\sim 20\%$.
(If only $pp$ and $^7$Be neutrinos are considered, the sum rule is
satisfied by construction in the one-zone model 
to the accuracy of our numerical
approximations.) 
Violations of the sum rule in the one-zone model 
are caused primarily by the fact that
different neutrino fluxes are produced in different temperature
regions of the sun.  If high precision is required, the solar neutrino
fluxes cannot be parameterized by a single temperature.

\section{Correlations Between Neutrino Fluxes}
\label{sec:correlations}

In the previous sections, we have concentrated (as have most other
investigations of this subject) 
on the average dependence of individual neutrino fluxes on the central
solar temperature.  In this section, we  focus on the
correlations that occur between the deviations of different neutrino
fluxes from their average values.  Previously we asked:  On average, 
how strongly does a particular neutrino flux depend upon temperature?
In this section, we ask:  If one neutrino flux is larger than its
average value by a specified amount, is a second neutrino flux larger
or smaller than its average and, if so, by how much? 

The one-zone model predicts the 
relative magnitude and the relative 
phase of the fractional changes, $\Delta
\phi/\phi$, of the $^7$Be and the $pp$ neutrino fluxes.  
Here 

\begin{equation}
\label{deltaphidef}
\Delta \phi_i = \phi_i -\phi_{i,{\rm SSM}},
\end{equation}
where
$\phi_{i,{\rm SSM}}$
is the standard value of the $i$th neutrino flux computed for the best
input parameters and input physics.
Since the
temperature dependence of both the $^7$Be and the $pp$ neutrino fluxes
are, in the approximation in which we are working, 
governed by the rate, $R_{34}$, of the ${\rm ^3He} + {\rm ^4He}$
reaction, the fractional changes in the fluxes are expected to be
proportional to each other.  The proportionality constant can be derived
by comparing Eq.~(\ref{ppnucrate}), Eq.~(\ref{ppfluxlum}), and
Eq.~(\ref{be7rate}).  We find

\begin{equation}
\label{be7proptopp}
{\Delta \phi({\rm ^7Be}) \over \phi({\rm ^7Be})_{\rm SSM} } ~=~
-
\left[ { \phi(pp)_{\rm SSM} \over \phi({\rm ^7Be})_{\rm SSM} } ~+~1 
\right]
{\Delta \phi(pp) \over \phi(pp)_{\rm SSM}}. 
\end{equation}

Figure~\ref{correlationbe7pp} shows, as the one-zone model predicts,
that the slope of the $\Delta \phi({\rm ^7Be})/\phi({\rm ^7Be})$ versus 
$\Delta \phi({\rm pp})/\phi({\rm pp})$ relation is negative and the
magnitude of the slope is
$ \sim -10$.  A closer study of
Figure~\ref{correlationbe7pp}
reveals  that the slope, $\alpha$, obtained with the 1000
numerical  models used in
the Monte Carlo study is ${\alpha}_{\rm Monte~Carlo} \approx -9$,
whereas the slope predicted by the one-zone solar model is
${\alpha}_{\rm one-zone} \approx -13$.  The difference between the
slope obtained with the Monte Carlo study and the slope found with the
one-zone model reflects the same imprecision in the one-zone model
that was found earlier in Section~\ref{subsec:ppTscaling} and 
Section~\ref{subsec:be7b8Tscaling}.  The slope that is relevant for
Figure~\ref{correlationbe7pp} is the ratio of the $^7$Be and $pp$
temperature exponents for neutrino fluxes; these exponents are
predicted by the one-zone model to be (see Table~\ref{tableexp}), 
respectively, 10\% too large
and 20\% too small relative to the detailed models.

The one-zone model also predicts the average linear relation between the
fractional changes, $\Delta \phi/\phi$, of the $^8$B, $^7$Be, and $pp$
neutrino fluxes.  
Figure~\ref{correlationb8be7pp} shows, in the top panel, the
fractional changes in flux for $^8$B neutrinos versus the fractional
changes in flux for the $^7$Be neutrinos,  for the 1000 detailed
solar models.  The bottom panel of Figure~\ref{correlationb8be7pp}
shows fractional changes in flux for $^8$B neutrinos versus $^7$Be
neutrinos.

The relevant slope in a plot of 
$\Delta \phi_i/\phi_i$ versus 
$\Delta \phi_j/\phi_j$ is just the ratio of the corresponding
temperature exponents for $\phi_i$ and $\phi_j$
that are given in Table~\ref{tableexp}. Therefore,
the predicted one-zone model slope, $\alpha$, for $^8$B versus $^7$Be is
${\alpha}_{\rm one-zone} \approx -2.3$,which is 
very close to the value of ${\alpha}_{\rm Monte~Carlo} \approx
-2.4$ found in the Monte Carlo study.
The one-zone model predicts a somewhat too steep dependence of
fractional changes in $^8$B
neutrino fluxes versus fractional changes in $^7$Be neutrino fluxes,
namely, ${\alpha}_{\rm one-zone} \approx 28$, versus 
${\alpha}_{\rm Monte~Carlo} \approx
22$.

Figure~\ref{correlationb8be7pp} shows a large scatter in 
the relation between
fractional changes of the $^8$B neutrino flux and either the 
$^7$Be or the $pp$ neutrino flux.  This lack of tightness in the
relations shown in Figure~\ref{correlationb8be7pp}
results ultimately from the fact that, in all modern solar models,
the $^8$B-producing reaction, 
reaction~8
Table~\ref{ppreaction}, is 
rare and does not influence significantly the 
structure of the sun. In fact, the largest uncertainty in the model
calculations of the $^8$B neutrino flux is caused by the 
uncertainty in the experimental value for the 
low-energy nuclear cross section of reaction~8; the value
of this cross section has essentially no effect ($ < 0.1 \%$) 
on the calculated rates of the other nuclear fusion reactions.      

\section{Conclusion and Discussion}
\label{sec:conclusion}

The most interesting result of our study is the understanding it
provides of the
negative temperature dependence of the $pp$ neutrino flux.  The
empirical fact that the $pp$ flux decreases with increasing central
temperature, contrary to the trend found with all other solar neutrino
fluxes, has been known since 1988\cite{bah88,bah89}, but has not
previously been explained physically.  At first glance, 
this negative temperature
dependence is counter-intuitive. 

In Section~\ref{subsec:ppTscaling}, 
we show that
the negative temperature dependence is a simple 
consequence of the fact that at higher temperatures only one $pp$
neutrino is produced per (approximately) 
$25$ MeV communicated to the star(via 
fusion of four protons), whereas at
lower temperatures two $pp$ neutrinos are produced per $25$ MeV.  
In other words, at lower temperatures the $^3$He-$^3$He fusion
termination reaction(which
requires two $pp$ reactions) predominates whereas at higher
temperatures the $^3$He-$^4$He reaction is faster (and requires only
one $pp$ reaction).  The total energy per unit time communicated 
to the star must equal the observed solar luminosity, 
independent of the assumed central temperature. Thus,  as the
temperature increases and  more of the
nuclear fusion is accomplished by the $^3$He-$^4$He reaction, fewer
$pp$ neutrinos are produced (and more $^7$Be and $^8$B neutrinos are
created).  

In order to obtain a simple physical understanding of the temperature
scalings and the correlations between the different neutrino fluxes,
we have adopted a one-zone model for the interior of the sun.  This
model is characterized by a fixed central temperature, $T_c$,
and a total luminosity that is equal to the observed solar luminosity.
Given the emphasis in the current literature on calculating ever more
precise solar models, with hundreds of different mass shells, it is
gratifying and surprising that the one-zone  model accounts
semi-quantitatively for some of the most often used 
results of the detailed model calculations. Moreover, the one-zone
model predicts the average correlations found between the different
neutrino fluxes, a bonus in insight that was not possible to
anticipate without detailed study of the simple model.

Figure~\ref{functions} and Figure~\ref{powerlaw} display the
dependence upon central 
temperature of the 1000 detailed solar models used in
the Bahcall-Ulrich
Monte Carlo study\cite{bah88,bah89} of theoretical uncertainties in
the predicted solar neutrino fluxes.  We have used these data 
to determine average temperature exponents, $m$, for all of 
the solar neutrino
fluxes, where by assumption $\phi \propto T^m$.  
The exponents determined here are obtained by  a formal best-fitting
technique and are to be preferred to the
previously-estimated exponents\cite{bah88,bah89} inferred less
formally from these same data; the previously-estimated exponents
have been widely used in the literature.
We have also
estimated, for the first time so far as we know, uncertainties in the
inferred temperature exponents.  

Figure~\ref{correlationbe7pp} and Figure~\ref{correlationb8be7pp} show
the correlations, found in the Monte Carlo study,
between the different individual neutrino fluxes.
These correlations reflect the fact that when one neutrino flux is
increased or decreased, there is likely to be a corresponding change
in the values of the other neutrino fluxes.  These correlations must
be taken into account when comparing the results of theoretical solar
model 
calculations, including their uncertainties, with solar neutrino
experiments. 
The only precise way to include the correlations displayed in 
Figure~\ref{correlationbe7pp} and Figure~\ref{correlationb8be7pp}
is to use the complete set of calculated neutrino fluxes in the
theoretical analysis(cf. \cite{haxton89}).  
Various practical approximations to this rather
cumbersome method have been discussed in the literature (see, for
example, \cite{w93,h94,fol95}).

The temperature exponents calculated with the aid of the one-zone
model agree with the exponents inferred from the Monte Carlo study of
precise solar models to an accuracy of 20\% or better for the three
most important solar neutrino fluxes: $pp$, $^7$Be, and $^8$B.  The
results are shown in Table~\ref{tableexp}, which compares the
exponents calculated with the one-zone model with
the results obtained from the detailed solar models.
Figure~\ref{scalingvstemp} shows that the scaling exponents calculated
in the one-zone solar model are not strongly dependent upon the
assumed characteristic central temperature, $T_c$ (taken here to be $T_c =
14\times 10^6$K).

The quantitative agreement between the results of the one-zone model
and the detailed models is impressive given the fact that the
temperature exponents vary from $m \sim -1$ for $pp$ neutrinos to 
$m \sim +24$ for $^8$B neutrinos.

The physical insight provided by the one-zone model suggests a new
form for the temperature dependence of the $pp$ neutrino flux, which
is given in Eq.~(\ref{preferred}).  In this form, the variation of the
$pp$ neutrino flux is, for all temperatures, consistent with the
observed solar luminosity, since it was derived by considering the
relation between the solar luminosity, Eq. (2), and the $pp$
neutrino flux, Eq. (3).  Moreover, the 
formula for the $pp$ neutrino flux, Eq.~(\ref{ppfluxlum}),
provided by the
one-zone model makes explicit the close correlation between the $^7$Be
and $pp$ neutrino fluxes that is manifest in
Figure~\ref{correlationbe7pp}.  
The expression  used in this paper for the $pp$ neutrino flux,
Eq.~(\ref{preferred}), was derived
by considering the relation between the solar luminosity,
Eq.~(\ref{L}),
and the $pp$ rate, Eq.~(\ref{ppnucrate}).
Physically, the strong correlation exists because the $^7$Be neutrino
flux is proportional to the rate of reaction 5 
of  Table~\ref{ppreaction} and
the $pp$ neutrino flux is proportional to a constant minus the rate of
reaction 5 (if we neglect the small contribution from CNO neutrinos).  

The one-zone model also accounts quantitatively for the average
correlation, shown in Figure~\ref{correlationb8be7pp},
between the $^8$B and $^7$Be neutrino fluxes, and between the
$^8$B and $pp$ neutrino fluxes.  

No simple model can, however, account
in detail for the scatter in the correlation plots shown in
Figure~\ref{correlationbe7pp} and Figure~\ref{correlationb8be7pp}.
The Monte Carlo experiments simulate  uncertainties in many
different parameters; the power-law fits in the figures represent only
the average response of the neutrino fluxes to the changes in all the
individual  parameters.
For analyses requiring a precise assessment of the correlations
between the different neutrino fluxes, a Monte Carlo study of
detailed solar models is required.

What have we learned from this study?  Improved temperature exponents
for the neutrino fluxes are now available, with estimates for the
uncertainties in the exponents.  A static one-zone model of
the sun accounts for the essential features of the temperature scaling
of the neutrino fluxes and even describes well the average
correlations between the fluxes.  The model does not provide a precise
description of the temperature dependences nor of the correlations
between the different fluxes.  
The exponents derived from the one-zone mode model do not satify
precisely the sum rule derived from the measured solar luminosity.

The fundamental reason that the one-zone model does not account
accuractely for all of the known results is that in precise solar
models each  neutrino flux is  produced in a different range of
temperatures.  One cannot represent the results of different
temperature ranges by a single parameter, $T_c$.

In the future, a new Monte Carlo study must be undertaken to determine
the temperature scalings and the correlations between the neutrino
fluxes when, as required by helioseismological measurements\cite{bah92},
diffusion is taken into account in the solar model calculations.  The
analysis of Bludman\cite{bludman96} suggests that the effects of
diffusion may alter the inferred temperature exponents by a
non-negligible amount when compared to the values given in this paper,
which are obtained from detailed solar models that do not include
diffusion\cite{bah88}. A Monte Carlo study is now underway that will
create 1000 solar models that include diffusion and other recent
refinements of the stellar model\cite{BPfuture}.

\acknowledgments
AU is supported in part by an NSF Graduate Research Fellowship.
The research of JNB is supported in part by NSF grant PHY-92-45317 
with the Institute for Advanced Study.
We are grateful to S. Bludman, N. Hata, W. Haxton, 
E. Lisi, P. Krastev, and P.
Langacker for valuable discussions and communications.

\begin{figure}
\caption[]{The $pp$ and $pep$ Neutrino Fluxes as a Function of
Central Solar Temperature. The top panel shows the $pp$ neutrino flux
versus the central solar temperature (expressed in units of $10^6$K).
The lower panel shows the $pep$ neutrino fluxes versus central
temperature. 
The circles correspond to 1000 representations of the standard solar
model calculated in a precise Monte Carlo study of the uncertainties
in the standard model solar neutrino fluxes\cite{bah88}. 
The two plotted lines represent a range of acceptable fits to the 
numerical data, which correspond to the indicated power-law
dependences upon central temperature.
The functional form for the $pp$ reaction,
$1 - 0.08 (T/T_{{\rm SSM}}^{n^\prime}$) is discussed in the text.}
\label{functions}
\end{figure}

\begin{figure}
\caption[]{The $^7$Be, $^8$B, $^{13}$N, and $^{15}$O Neutrino Fluxes
as a Function of Central Solar Temperature. The four panels show how
the four different neutrino fluxes depend upon central temperature.
The circles correspond to 1000 numerical  solar models that were
computed with different
input data\cite{bah88}.
The plotted lines represent a range of acceptable fits to the
numerical data, which correspond to the indicated power-law
dependences upon temperature.}
\label{powerlaw}
\end{figure}

\begin{figure}
\caption[]{The Power-Law Exponents.  The figure shows how the
calculated power-law exponents($m$, where 
neutrino flux $\propto T_c^m$) depend
upon the characteristic central temperature of the one-zone model.
For example, $m(pp)$ varies between $-0.9$ to $-0.8$ as the central
temperature is varied between $12
\times 10^6$K to $16 \times 10^6$K. }
\label{scalingvstemp}
\end{figure}

\begin{figure}
\caption[]{Correlation of the $^7$Be and the $pp$ neutrino fluxes. The
figure shows the strong correlation, predicted by the one-zone model, 
between the fractional changes in the $^7$Be
neutrino flux and the fractional changes in the $pp$ neutrino flux as
calculated in the 1000 numerical  solar models  in the Monte Carlo
experiment of reference\cite{bah88}.
Here $\Delta \phi_i = \phi_i -\phi_{i,{\rm SSM}}$.}
\label{correlationbe7pp}
\end{figure}

\begin{figure}
\caption[]{Correlations between the $^7$Be, $^8$B, and $pp$ neutrino
fluxes. The figure shows the moderate correlation that exists between
the fractional changes in the $^8$B neutrino flux and the fractional
changes of either the $^7$Be or the $pp$ neutrino flux.
Here $\Delta \phi_i = \phi_i -\phi_{i,{\rm SSM}}$.}
\label{correlationb8be7pp}
\end{figure}

%\begin{figure}[f1]
%\plotone{composite3.ps}
% \caption{}
%\end{figure}

%\begin{figure}[f2]
%\plotone{composite2.ps}
% \caption{}
%\end{figure}

%\begin{figure}[f3]
%\plotone{fig9.ps}
% \caption{}
%\end{figure}

\clearpage

\begin{table}[htb]
\tightenlines
\mediumtext
\centering
\begin{minipage}[htb]{4.1in}
\caption[]{The Principal Reactions of the $pp$ Chain\label{ppreaction}}
\begin{tabular}{lcc}
\multicolumn{1}{c}{Reaction}&Reaction&Neutrino Energy\\
&Number&(MeV)\\
\noalign{\smallskip}
\hline
\noalign{\medskip}
$\phantom{^3}p + p \to {\rm ^2H} + e^+ + \nu_e$&1&0.0 to 0.4 \\
\ \ \ \ $\phantom{^8}p + e^- + p \to {\rm ^2H} + \nu_e$&2&1.4 \\
${\rm ^2H} + p \to {\rm ^3He} + \gamma$&3 \\
${\rm ^3He} + {\rm ^3He} \to {\rm ^4He} + 2p$&4 \\
\multicolumn{1}{c}{or} \\
${\rm ^3He} + {\rm ^4He} \to {\rm ^7Be} + \gamma$&5 \\
\multicolumn{1}{c}{then} \\
\ \ \ \ $\phantom{^3}e^- + {\rm ^7Be} \to {\rm ^7Li} + \nu_e$&6&0.86,
0.38 \\
\ \ \ \ \ \ \ \ ${\rm ^7Li} + p \to {\rm ^4He} + {\rm ^4He}$&7 \\
\multicolumn{1}{c}{or} \\
\ \ \ \ $\phantom{^3}p + {\rm ^7Be} \to {\rm ^8B} + \gamma$&8 \\
\ \ \ \ \ \ \ \ ${\rm ^8B} \to {\rm ^8Be} + e^+ + \nu_e$&9&0 to 14 \\
\end{tabular}
\smallskip
\end{minipage}
\end{table}

\bigskip
\begin{table}[htb]
\tightenlines
\mediumtext
\centering
\begin{minipage}[htb]{5in}
\caption[]{Temperature exponents for solar neutrino fluxes. 
We recommend a new functional dependence for $\phi(pp)$,
$\phi(pp) \propto 1  - 0.08 (T/T_{{\rm SSM}})^{m^\prime}$
as discussed in the text.
All other exponents are given for the functional form, $\phi \propto T^m$.
For the one-zone model, we assumed a characteristic temperature
$T_c = 14 \times 10^6$K.
\label{tableexp}}
\begin{tabular}{lccc}
Neutrino &Monte Carlo
&Estimated
&One-Zone  \\
Flux &Exponent &
Uncertainty &Exponent  \\
\hline
$\phi(pp)$,$m^\prime  $  &    13.0     &     0.7      	& 11 \\
$\phi(pp)$,$m         $  &    \llap{$-$}1.1     &     0.1      	& 
\llap{$-$}0.9 \\
$\phi(pep)$ &    \llap{$-$}2.4     &     0.9    	& \llap{$-$}1.4   \\        
$\phi(^7$Be) &     10     &     2		& 11 \\        
$\phi(^8$B)   &     24     &      5     	& 25 \\        
$\phi(^{13}$N)  &    24.4     &     0.2   	& 20   \\      
$\phi(^{15}$O)  &    27.1     &     0.1   	& 20   \\      
$\phi(^{17}$F)  &    27.8     &     0.1    	& 23  \\   
\end{tabular}
\end{minipage}
\end{table}

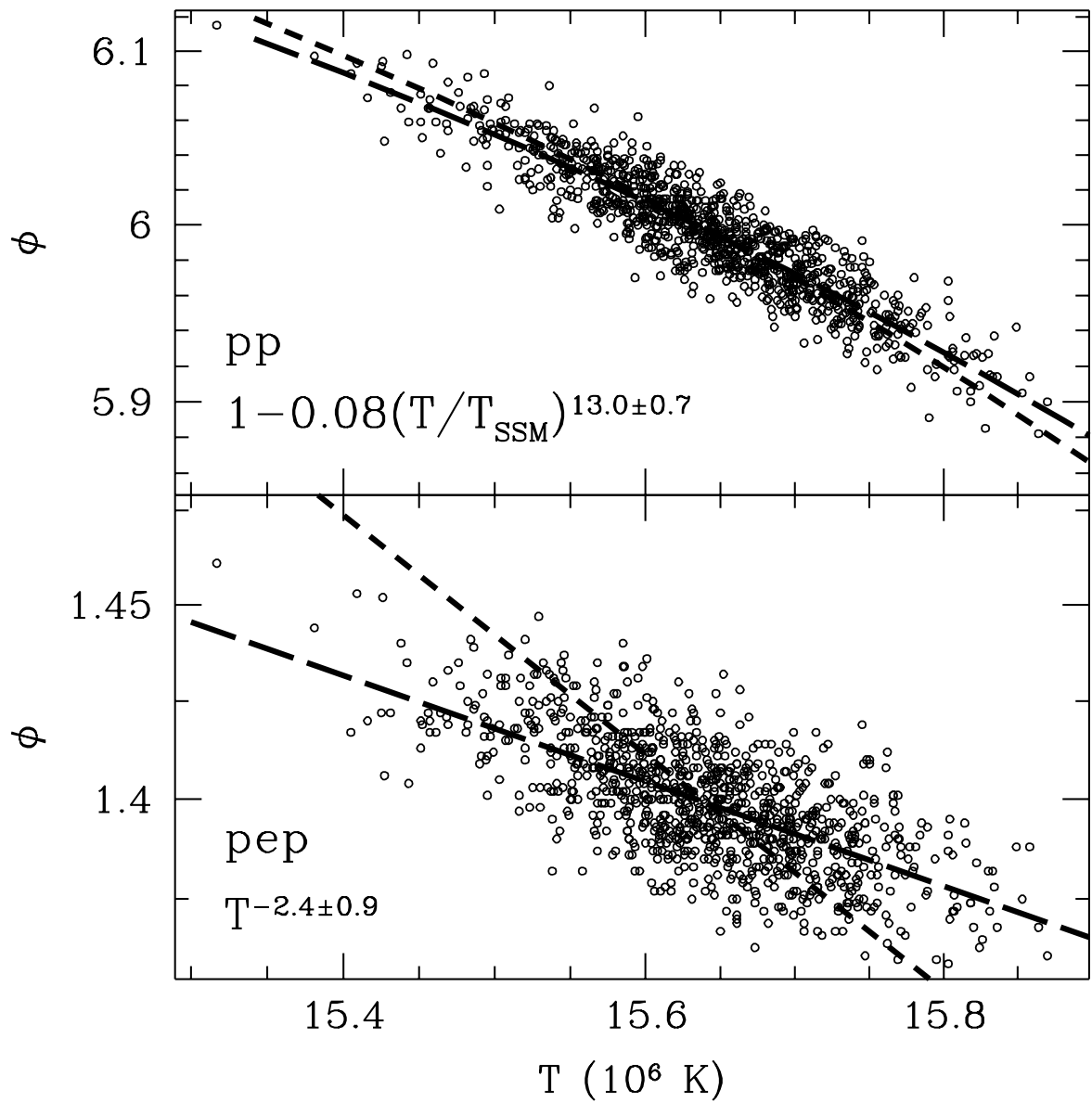
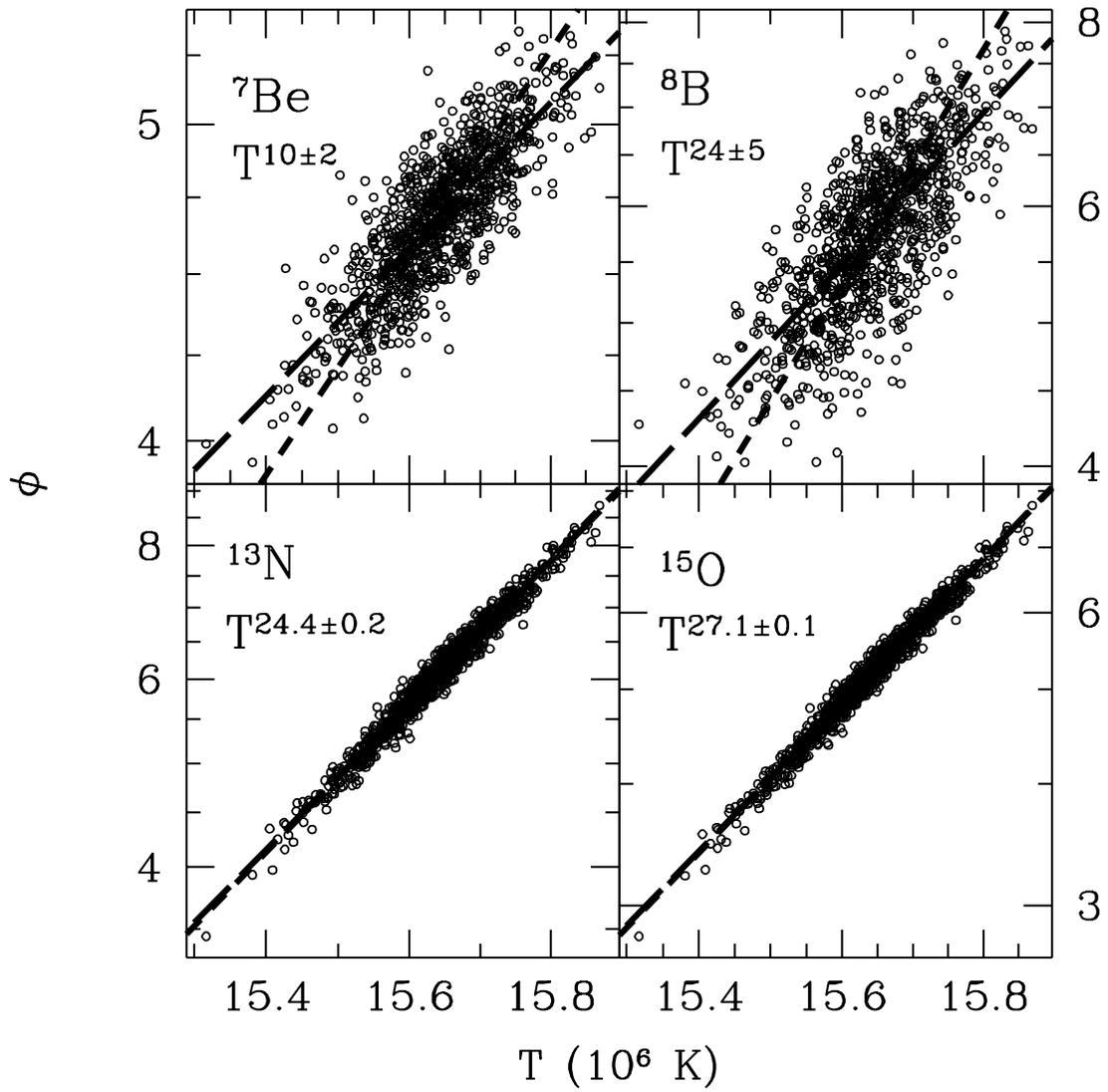
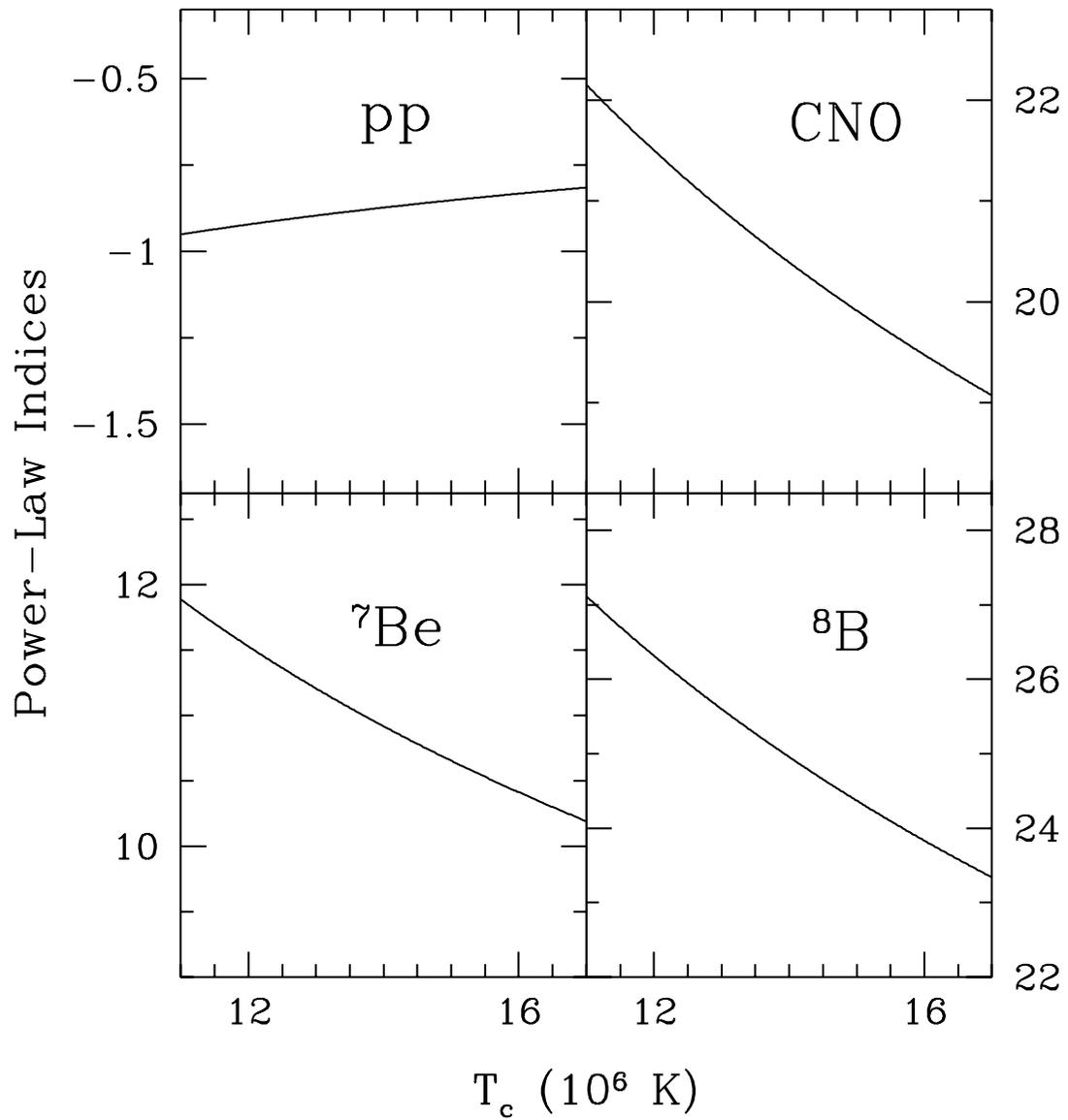
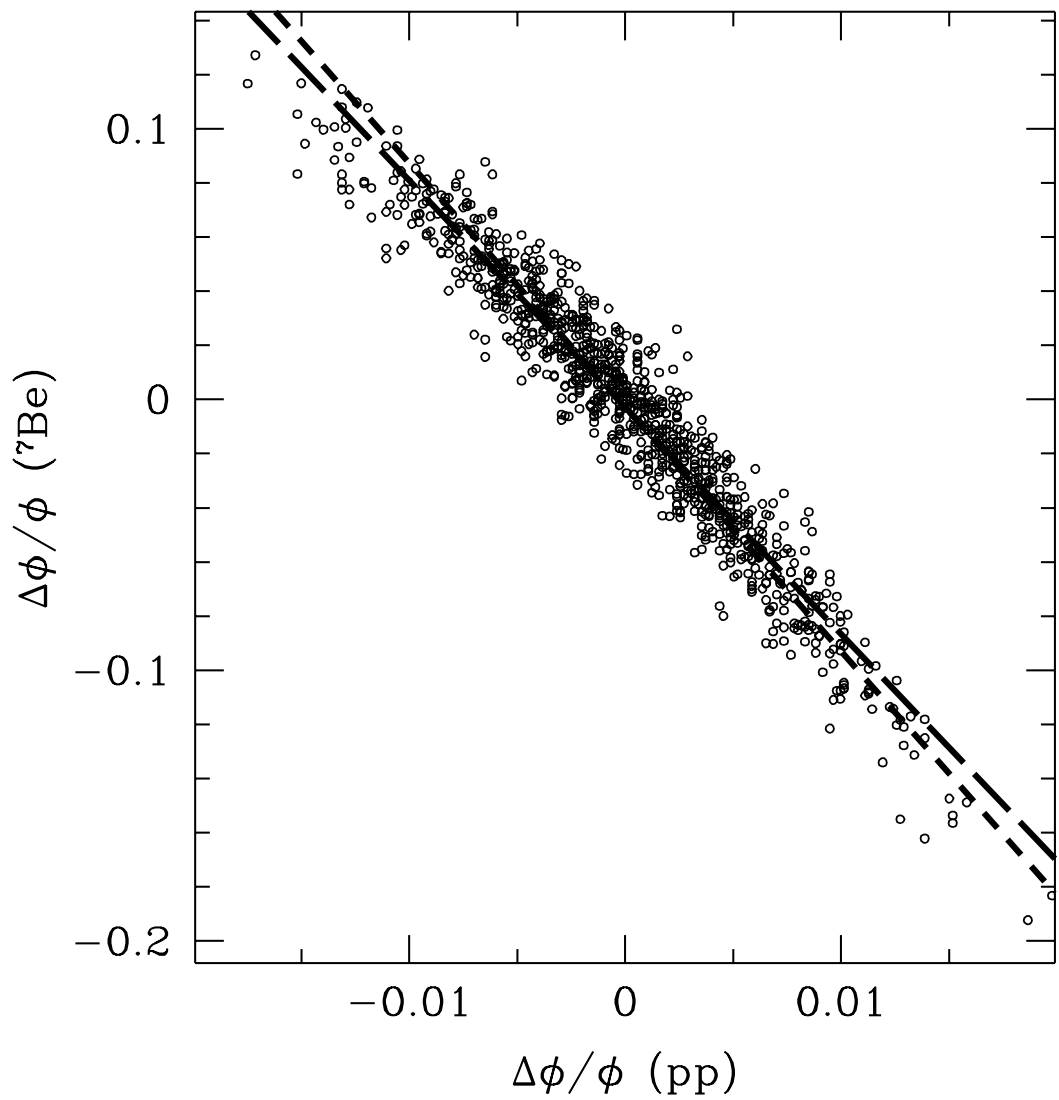
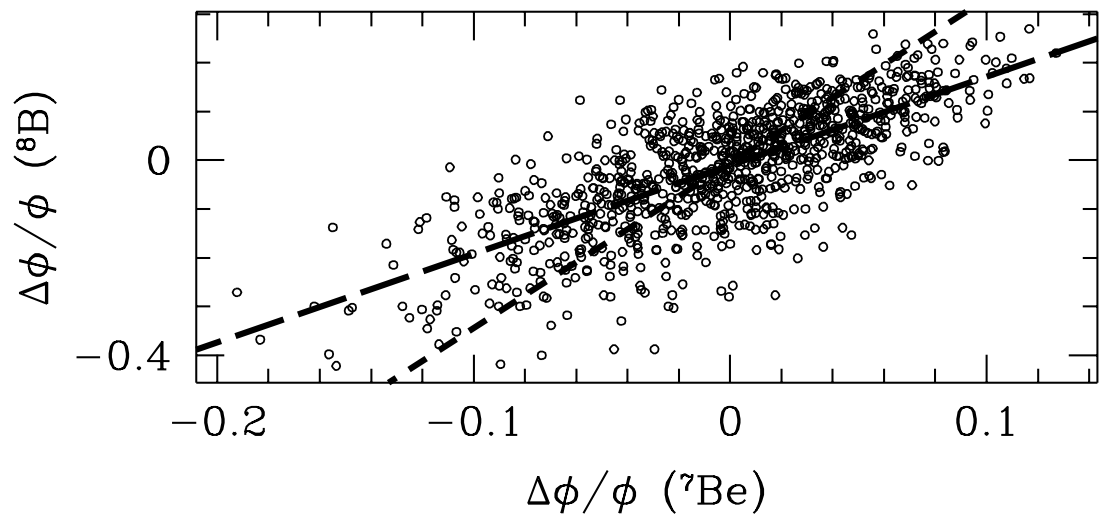
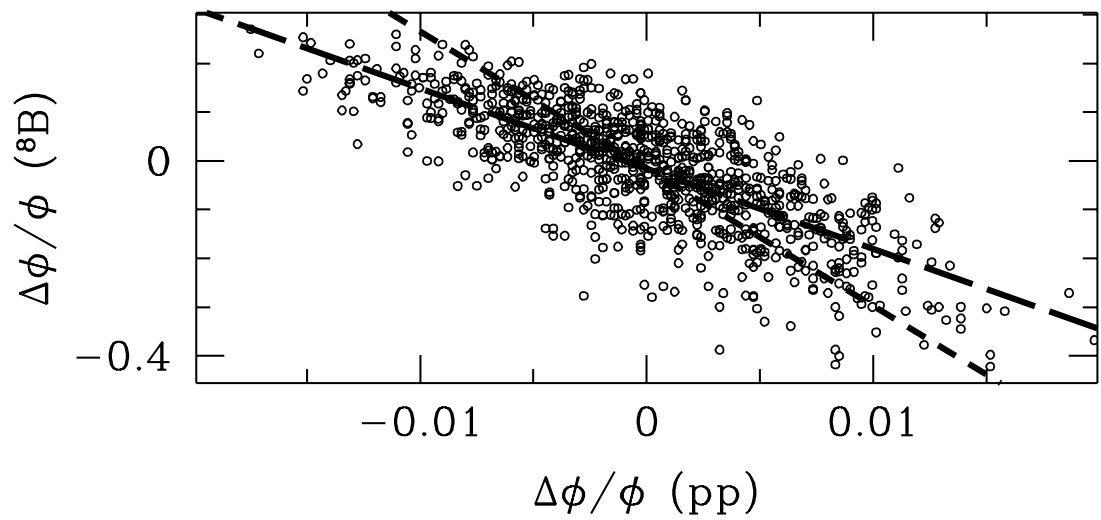
\end{document}